\documentclass[english,aps,manuscript]{revtex4}
\usepackage[T1]{fontenc}
\usepackage[latin9]{inputenc}
\usepackage{graphicx}
\usepackage{amssymb}

\providecommand{\tabularnewline}{\\}

\usepackage{babel}

\begin{document}

\title{Searching for Higgs Boson at the Tevatron: the {}``Golden Mode''
revisited}

\author{M. SAHIN}

\email{m.sahin@etu.edu.tr}

\affiliation{TOBB University of Economics and Technology, Physics Division, Ankara,
Turkey}

\author{S. SULTANSOY}

\email{ssultansoy@etu.edu.tr}

\affiliation{TOBB University of Economics and Technology, Physics Division, Ankara,
Turkey}

\affiliation{Institute of Physics, National Academy of Sciences, Baku, Azerbaijan}

\author{S. TURKOZ}

\email{turkoz@science.ankara.edu.tr}

\affiliation{Ankara University, Department of Physics, Ankara, Turkey}
\begin{abstract}
We have reconsidered the Higgs boson search via the {}``golden mode''
for Tevatron. It is shown that this mode will give opportunity to
observe the Higgs boson with mass up to $300$ GeV before Tevatron
shutdown. 
\end{abstract}
\maketitle
The existence of the fourth family is well motivated in the framework
of the standard model (SM) \cite{PMC,sahin}. According to flavor
democracy hypothesis (FDH) Dirac masses of the fourth family fermions
are approximately equal \cite{Fritzsch,Datta,Celikel}. If common
Yukawa coupling constant is taken to be equal to $g_{W}$, the Dirac
masses of the fourth SM family fermions are predicted as $m_{4}\simeq450$
GeV at $\mu\backsimeq245$ GeV (vacuum expection value of the neutral
component of the Higgs doublet) scale. Concerning pole masses we have
$m_{l_{4}}\thickapprox m_{\nu_{4}}(Dirac)\thickapprox450$ GeV, $m_{d_{4}}\backsimeq500$
GeV and $m_{u_{4}}\thickapprox m_{d_{4}}\pm50$ GeV. If the $\nu_{4}$
has a Majorana nature, the mass of the lighter one could be down to
$50$ GeV.

Actually, the fourth SM family quarks are favorite candidates for
the LHC discoveries after the Higgs boson (it is interesting that
if Higgs quartic coupling constant is equal to $g_{W}$ the mass of
the SM Higgs boson is expected to be around $290$ GeV). Moreover,
fourth family quarks strongly affect Higgs boson production at hadron
colliders: the gluon fusion channel is enhanced by a factor up to
$9$ for low values of the Higgs boson mass, and the enhancement factor
is reduced to minimal value of $4.5$ at $m_{H}\thickapprox500$ GeV
(for details see e.g. \cite{Becerici}). While this enhancement seems
almost obvious, it should be emphasized that concerning the LHC and
Tevatron this subject was considered well before 2007 \cite{EnginArik1,PJenni,Ginzburg,Sultansoy1,EnginArik2,EnginArik3,Cakir1,EnginArik4,Kribs1,EnginArik5}.

During the last two decades, the fourth SM family studies were almost
blocked by incorrect interpretation of the precision electroweak data.
Despite to the studies done 10 years ago \cite{Maltoni,He,Vysotsky,Novikov,Bulanov}
this misinterpretation continued to have a place in Particle Data
Group (PDG) up to 2008. Lately, authors of the corresponding part
of PDG have come \cite{PDG2010,Erler} close to common understanding
\cite{Maltoni,He,Vysotsky,Novikov,Bulanov,Novikov1,Alwall,Kribs1,Bobrowski,Chanowitz,Hashimoto,Ebberhardt,Cobanoglu,OPUCEM,sahin}.

For illustration by using OPUCEM software \cite{OPUCEM} we have shown
that SM4 points with above mentioned values for fourth family fermions
are in better agreement with precision electroweak data than SM3.
Corresponding points are presented in Table I where $m_{l_{4}}=450$
GeV, $m_{d_{4}}=500$ GeV, $s_{34}=0.01$ (Cabibbo-Kobayashi-Maskawa
mixing between fourth and third SM family quarks), $m_{\nu_{4}}(l)=110$
GeV for fourth family light Majorana neutrino and the rest are given
in the Table I. In Figure 1 we present these three points in the S-T
plane together with SM3 prediction for $m_{H}=115$ GeV. It is seen
that SM4 points are closer to central values of S and T parameters.
In Figure 2 the points corresponding to $m_{H}=200,\,250$ and $300$
GeV are presented for SM3 case. As it is seen these points are outside
$2\sigma$ counter. 

\begin{table}
\begin{tabular}{|c|c|c|c|c|}
\hline 
SM4 points  & 1  & 2  & 3  & SM3\tabularnewline
\hline
\hline 
$m_{u_{4}}$, GeV  & $540$  & $540$  & $520$  & -\tabularnewline
\hline 
$m_{\nu_{4}}$(h), GeV  & $2200$  & $2000$  & $1600$  & -\tabularnewline
\hline 
$m_{H}$, GeV  & $200$ & $250$  & $300$  & $115$\tabularnewline
\hline
\end{tabular}\caption{SM4 points for three different values of $m_{H}$.}

\end{table}

\begin{figure}
\includegraphics[scale=0.7]{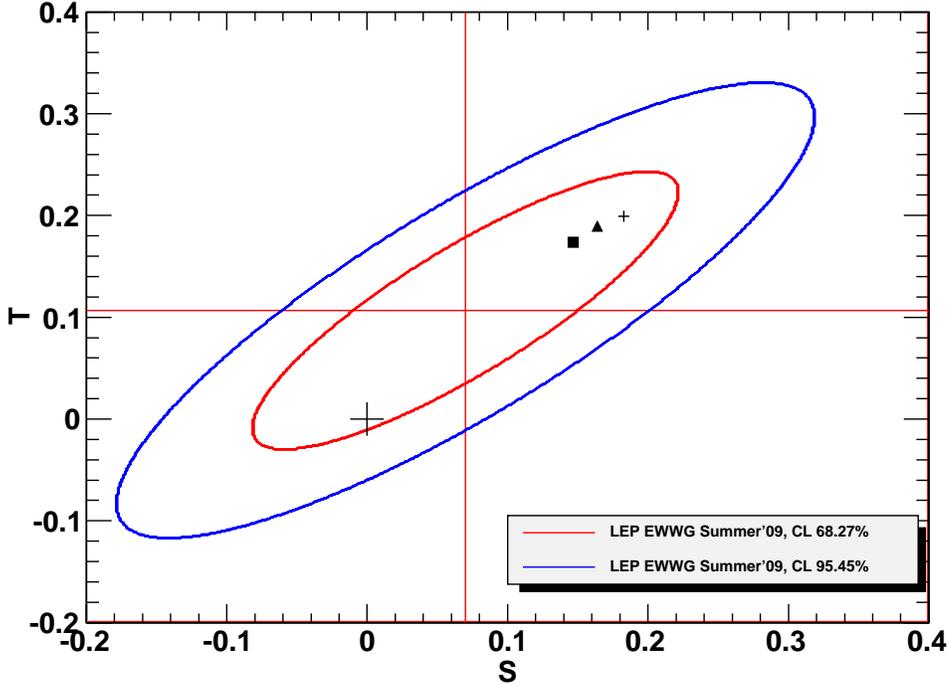}

\caption{SM3 and three SM4 points in S-T plane. The 1 and 2$\sigma$ error
ellipses represent the 2009 results of the U = 0 fit from LEP EWWG.
Large cross corresponds to SM3 with $m_{H}=115$ GeV; square, triangle
and small cross symbols correspond to SM4 points 1, 2 and 3 from Table
I, respectively. This figure is obtained by using OPUCEM software. }

\end{figure}

\begin{figure}
\includegraphics[scale=0.7]{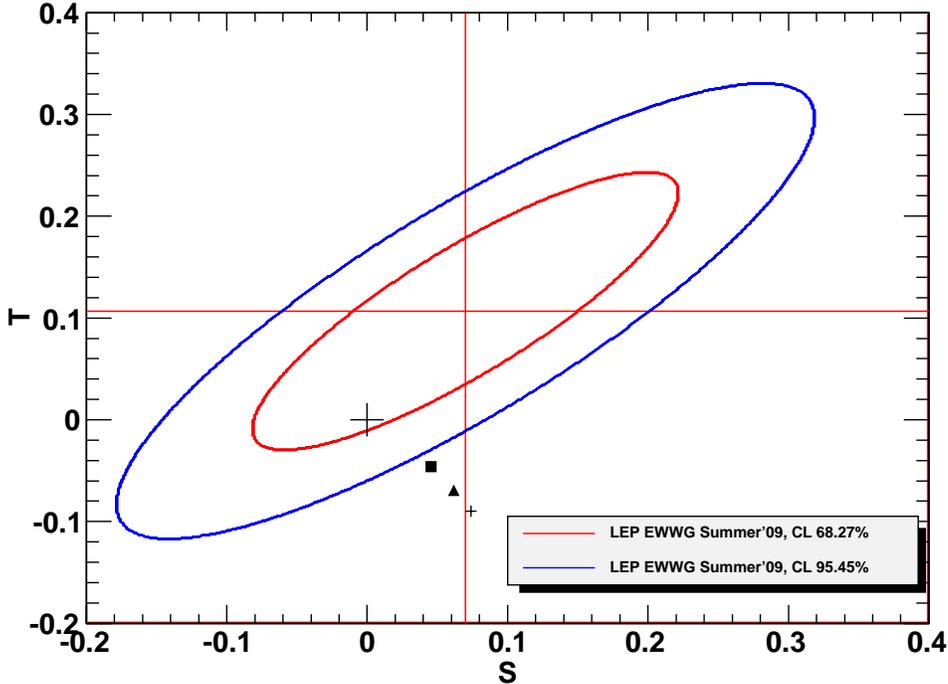}

\caption{SM3 points with $m_{H}=200$ (square) , $250$ (triangle) and $300$
(small cross) GeV in S-T plane. This figure is obtained by using OPUCEM
software. }

\end{figure}

Today, the race between the LHC and Tevatron on Higgs search, as well
as fourth SM family quarks searches, is been observed. This competition
will continue by autumn 2011 (the date of the final shutdown for the
Tevatron) and by this time the LHC will exceed the integrated luminosity
of $1fb^{-1}$. In the case of four SM families this luminosity will
give an opportunity to scan various $m_{H}$ ranges via gluon fusion
mode: $140<m_{H}<250$ GeV in $H\rightarrow W^{+}W^{-}$ channel,
$135<m_{H}<160$ GeV and $175<m_{H}<400$ GeV in {}``golden mode''
($H\rightarrow ZZ\rightarrow4l$) channel \cite{Becerici}, $110<m_{H}<140$
GeV in $\tau^{+}$$\tau^{-}$ channel \cite{Baglio} (the authors
considered SM3 case, their results which easily can be rescaled to
SM4 case by taken in to account of the enhancement factor $\thicksim9$),
$350<m_{H}<700$ GeV in $H\rightarrow ZZ$ channel following by $ZZ\rightarrow\nu\overline{\nu}l^{+}l^{-}$
or $ZZ\rightarrow\nu\overline{\nu}q\overline{q}$ decays.

In this work we have reconsidered the {}``golden mode'' for the
Tevatron in the presence of the fourth SM family, proposed 10 years
ago \cite{Sultansoy1,Cakir1}. At that time it was expected that LHC
with $\sqrt{s}=14$ TeV would come to operation in 2005. Therefore,
further work on the subject have not been done. In this paper we investigate
the opportunity to observe Higgs boson via {}``golden mode'' at
the Tevatron with 20 $fb^{-1}$ integrated luminosity which correspond
to combination of D0 and CDF data that will be obtained before the
shutdown.

Up to now the search for the Higgs boson at the Tevatron in the case
of four SM families was concentrated on $gg\rightarrow H\rightarrow W^{+}W^{-}$
channel \cite{Abazov,Abulencia}. The latest combined results \cite{Aaltonen},
based on $L_{int}=4.8fb^{-1}$ at CDF and $L_{int}=5.4fb^{-1}$ at
D0, excluded Higgs boson with a mass between $131$ and $204$ GeV
at the 95\% confidence level. In \cite{Sultansoy1,Cakir1} it was
shown that the {}``golden mode'' ($gg\rightarrow H\rightarrow ZZ\rightarrow4l$)
could be effective for $m_{H}\geq180$ GeV at the upgraded Tevatron.
However, in the analysis only the cut on invariant mass was used,
together with overestimating expression for statistical significance.

In order to determine appropriate cuts, we calculate $p_{T}$ distributions
for final state leptons and $Z$ bosons, eta ($\eta$) distributions
for final state leptons, as well as four-lepton invariant mass distributions
for signal and background. For latter we consider irreducible background
from pair production of $Z$ bosons. In numerical calculations PYTHIA
\cite{Torbjorn } and MCFM \cite{Campbell1,Campell2,Campbell3,Campbell4,Berger,Campbell5}
simulation programs were used. The results are presented in Figures
2-5. We will use $p_{T}^{l}>10$ GeV and $|\eta^{l}|<2$ as generic
detector cuts (effect of these cuts on signal and background are similar
and small). For the four lepton invariant mass we use $m_{4l}=m_{H}\pm10$
GeV, which reduces background essentially, whereas signal is almost
unchanged. Another effective cut is provided by $p_{T}^{Z}$. Looking
at Figure 6 we decided to use $p_{T}^{Z}>30$, $p_{T}^{Z}>70$ and
$p_{T}^{Z}>100$ GeV for $m_{H}=200,\,250$ and $300$ GeV, respectively.

\begin{figure}
\includegraphics[scale=0.7]{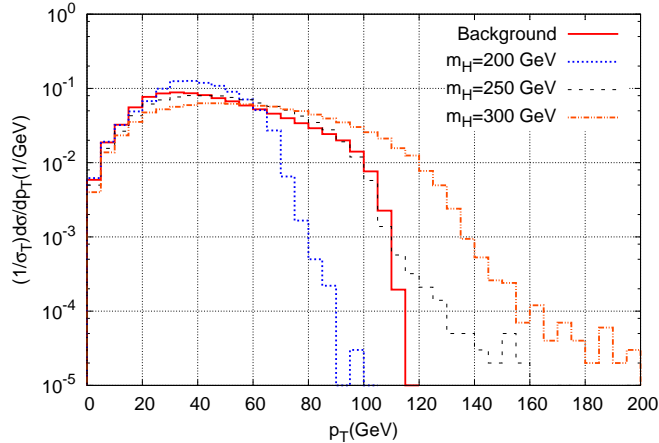}

\caption{Normalized $p_{T}^{l}$ distributions of final state leptons for signal
and background at the Tevatron. }

\end{figure}

\begin{figure}
\includegraphics[scale=0.7]{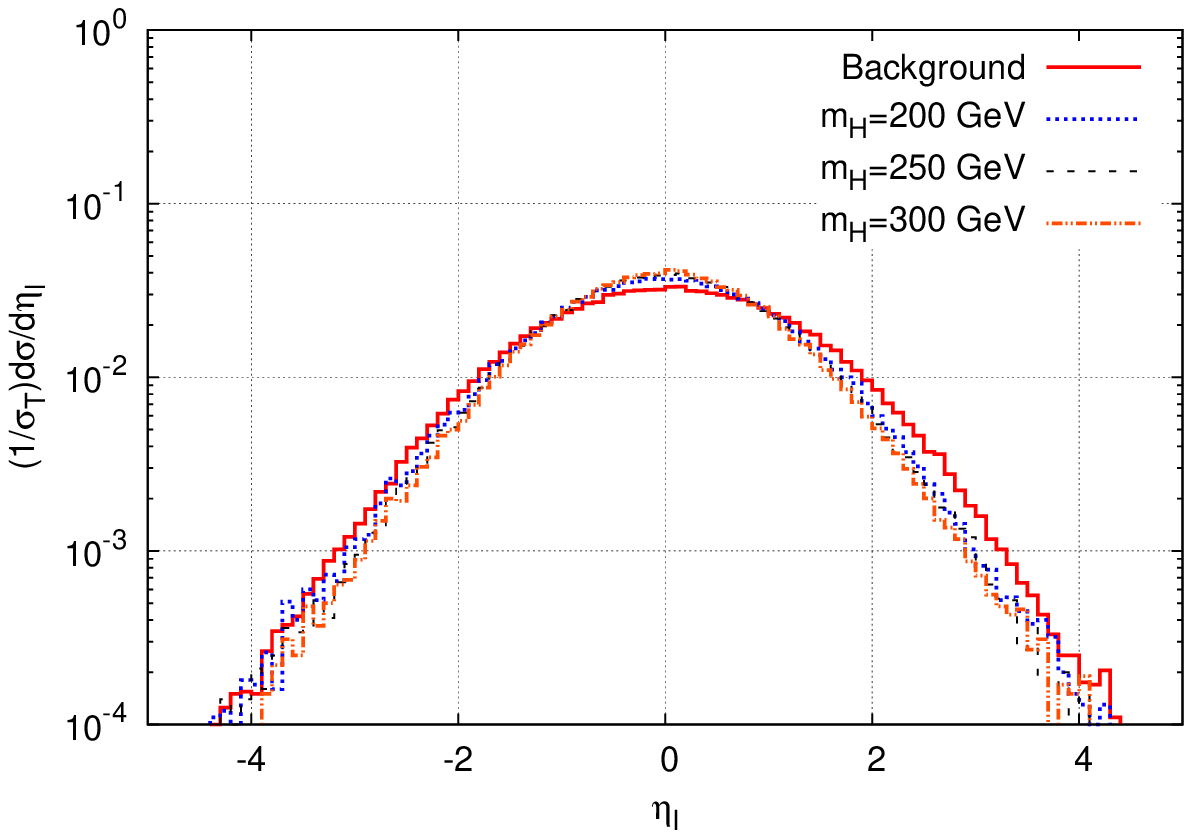}

\caption{Normalized $\eta^{l}$ distributions of final state leptons for signal
and background at the Tevatron.}

\end{figure}

\begin{figure}
\includegraphics[scale=0.7]{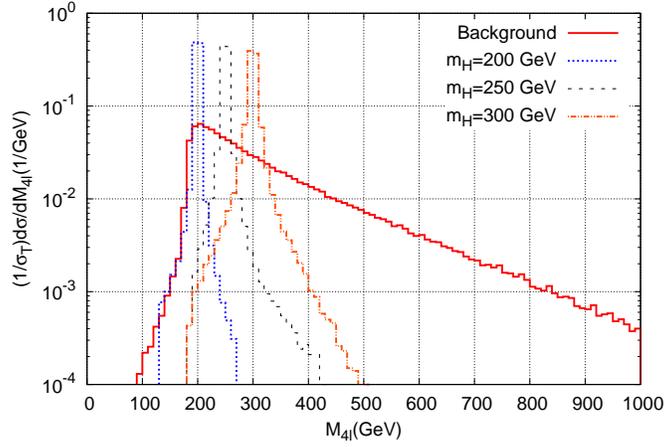}

\caption{Normalized invariant mass distributions of final state $4$ leptons
for signal and background at the Tevatron. }

\end{figure}

\begin{figure}
\includegraphics[scale=0.7]{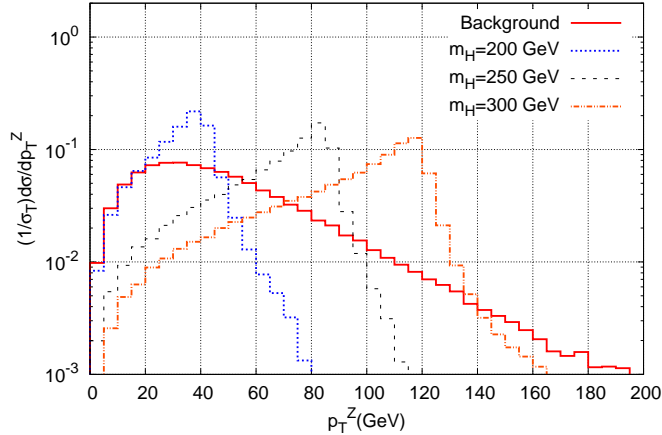}

\caption{Normalized $p_{T}^{Z}$ distributions of $Z$ bosons for signal and
background at the Tevatron}

\end{figure}

In Table II we present signal and background cross sections without
and with cuts for three different values of the Higgs mass. For $\sigma(H)$
without cuts, following Ref. \cite{Aaltonen}, we use results of NNLO
calculations from \cite{Charalampos}. 

\begin{table}
\begin{tabular}{|c|c|c|c|}
\hline 
$m_{H}$, GeV & $200$ & $250$ & $300$\tabularnewline
\hline
\hline 
$\sigma(H)$, w/o cuts & $1580$ & $650$ & $300$\tabularnewline
\hline 
$\sigma(H\rightarrow4l)$ w/o cuts & $1.789$  & $0.882$  & $0.421$ \tabularnewline
\hline 
$\sigma(H\rightarrow4l)$ with cuts & $1.186$  & $0.488$  & $0.183$\tabularnewline
\hline 
$\sigma(bkgr)$ w/o cuts & $4.364$  & $4.364$  & $4.364$ \tabularnewline
\hline 
$\sigma(bkgr)$ with cuts & $0.287$  & $0.104$  & $0.037$ \tabularnewline
\hline
\end{tabular}

\caption{Signal and background cross sections (in fb) without and with cuts. }

\end{table}

Statistical significance has been calculated by using following formula
\cite{CMS_Note_significance}:

\begin{equation}
S=\sqrt{2[(s+b)ln(1+\frac{s}{b})-s]}\end{equation}

where $b$ and $s$ represents the numbers of background and signal
events, respectively. In Table III, we present achievable statistical
significances for $L_{int}=10$ and $20$ $fb^{-1}$. Before the final
shutdown of the Tevatron each detector will reach approximately $L_{int}=10$
$fb^{-1}$integrated luminosity, which will allow to exclude $m_{H}$
up to $300$ GeV, whereas combination of results of two detectors
will give opportunity to observe Higgs boson via {}``golden mode''
if its mass is between $180$ and $300$ GeV. Comparing $W^{+}W^{-}$
mode analysis \cite{Aaltonen} which excludes $m_{H}$ between 134
and 204 GeV with the {}``golden mode'', we conclude that the same
amount of analyzing data (combined $L_{int}\simeq10$ $fb^{-1}$)
will give opportunity to discover Higgs boson if $m_{H}=200$ GeV,
or observe it if $m_{H}=250$ GeV. 

\begin{table}
\begin{tabular}{|c|c|c|c|}
\hline 
$m_{H}$, GeV & 200 & 250 & 300\tabularnewline
\hline
\hline 
$L_{int}=10$ $fb^{-1}$ & $4.9$$\sigma$ & $3.3$$\sigma$ & $2.0$$\sigma$\tabularnewline
\hline 
$L_{int}=20$ $fb^{-1}$ & $7.0$$\sigma$ & $4.7$$\sigma$ & $2.9$$\sigma$\tabularnewline
\hline
\end{tabular}

\caption{Achievable statistical significances. }

\end{table}

This work is supported by DPT and TUBITAK.

\end{document}